\documentclass[shortnote,preprint]{jpsj2}

\def\runauthor{Kazutaka {\sc Kudo} {\it et al.}}

\title{Multi-Triplet Magnons in SrCu$_2$(BO$_3$)$_2$ Studied by Thermal Conductivity Measurements in Magnetic Fields}

\author{Kazutaka \textsc{Kudo}\thanks{E-mail address: kudo@imr.tohoku.ac.jp}\thanks{Present address: Institute for Materials Research, Tohoku University, 2-1-1 Katahira, Aoba-ku, Sendai 980-8577}, Takashi \textsc{Noji}, 
Yoji \textsc{Koike}, Terukazu \textsc{Nishizaki}$^1$ and Norio \textsc{Kobayashi}$^1$
}

\inst{%
Department of Applied Physics, Graduate School of Engineering, Tohoku University, 6-6-05 Aoba Aramaki, Aoba-ku, Sendai 980-8579 \\ 
$^1$Institute for Materials Research, Tohoku University, 2-1-1 Katahira, Aoba-ku, Sendai 980-8577 \\
}

\recdate{\today}

\kword
{
thermal conductivity, magnon, spin gap
}

\begin{document}
\sloppy
\maketitle

Recently, thermal conductivity in low-dimensional quantum spin systems has attracted interest, because the thermal conductivity due to magnons, $\kappa_{\rm magnons}$, has been found to be large in some low-dimensional copper-oxides\cite{rf:Kudo3M}. 
Theoretically, the existence of a ballistic $\kappa_{\rm magnon}$ has been predicted in integrable systems\cite{rf:Saito}, and actually a large thermal conductivity associated with the ballistic $\kappa_{\rm magnon}$ has been observed in one-dimensional Heisenberg chain systems SrCuO$_2$ and Sr$_2$CuO$_3$\cite{rf:Solo112}. 
As for non-integrable systems such as spin-gap systems, on the other hand, $\kappa_{\rm magnon}$ has theoretically been predicted to be diffusive and small. 
Experimentally, however, a large $\kappa_{\rm magnon}$ has been observed in a spin-gap state of the two-leg spin-ladder system Sr$_{14}$Cu$_{24}$O$_{41}$\cite{rf:KudoLT,rf:Solo,rf:Kudo1,rf:Kudo2,rf:Hess}, owing to the marked decrease of the magnon-magnon scattering. 
Therefore, the mechanism of large $\kappa_{\rm magnon}$ has not yet been understood so clearly.

The compound SrCu$_2$(BO$_3$)$_2$ is a layered material composed of the two-dimensional CuBO$_3$ plane, as shown in the inset of Fig. \ref{fig:kappa}(a). 
The ground state is a spin-singlet state with the spin gap of 34 K\cite{rf:Kageyama1,rf:plataeu2,rf:neutron} and can be expressed in terms of the Shastry-Sutherland model\cite{rf:Kageyama1,rf:Miyahara,rf:Shastry}. 
In the spin gap state of SrCu$_2$(BO$_3$)$_2$, it has been reported that the thermal conductivity due to phonons, $\kappa_{\rm phonon}$, is enhanced, owing to the decrease of the phonon-magnon scattering, and that magnons do not contribute to the thermal conductivity\cite{rf:KudoSrCu2,rf:HoffSrCu2,rf:KudoSrZn}. 
This is because single-triplet magnons in the first-excited state are extremely localized, according to the magnetization curve measurements\cite{rf:Kageyama1,rf:plataeu2}, the inelastic neutron scattering experiment\cite{rf:neutron} and the perturbation calculations\cite{rf:Miyahara}. 
On the other hand, the theoretical calculation has suggested that the excited state with higher energy composed of two or several triplet magnons (multi-triplet magnons) can propagate rather fast\cite{rf:Momoi1,rf:Momoi2}.  
In fact, the existence of such multi-triplet magnons has been confirmed experimentally in the ESR\cite{rf:ESR}, the inelastic neutron scattering\cite{rf:neutron}, the far-infrared absorbtion\cite{rf:far} and the Raman scattering\cite{rf:Raman} measurements.
The multi-triplet magnons may contribute to the thermal conductivity when the energy bands are lowered by the application of magnetic field.  
In order to distinguish the contributions of $\kappa_{\rm magnon}$ and $\kappa_{\rm phonon}$ from each other, it is effective to study effects of the partial substitution of Zn for Cu on the thermal conductivity, because Zn$^{2+}$(the spin quantum number $S =$ 0) is expected to shorten the mean free path of magnons running in the Cu$^{2+}(S = 1/2)$ network\cite{rf:Kudo2}. 
In this paper, we have measured the thermal conductivity parallel to the $a$-axis (i.e. parallel to CuBO$_3$ plane)  of the Zn-free and 1\% Zn-substituted SrCu$_{2-x}$Zn$_x$(BO$_3$)$_2$ in magnetic fields up to 14 T, in order to examine $\kappa_{\rm magnon}$ due to the multi-triplet magnons. 
We have also measured the magnetic susceptibility to check the magnetic properties.

Single crystals of SrCu$_{2-x}$Zn$_{x}$(BO$_3$)$_2$ with $x =$ 0 and 0.02 were grown using the traveling-solvent floating-zone method\cite{rf:KudoSrCu2,rf:single}. 
Thermal conductivity measurements were carried out using the conventional steady-state technique. 
Magnetic fields up to 14 T were applied parallel to the thermal-current direction, using a superconducting magnet. 
The magnetic susceptibility was measured in a magnetic field of 1 T using a SQUID magnetometer (Quantum Design, MPMS-XL5).

\begin{figure}[p]
\begin{center}
\includegraphics[width=0.5\linewidth]{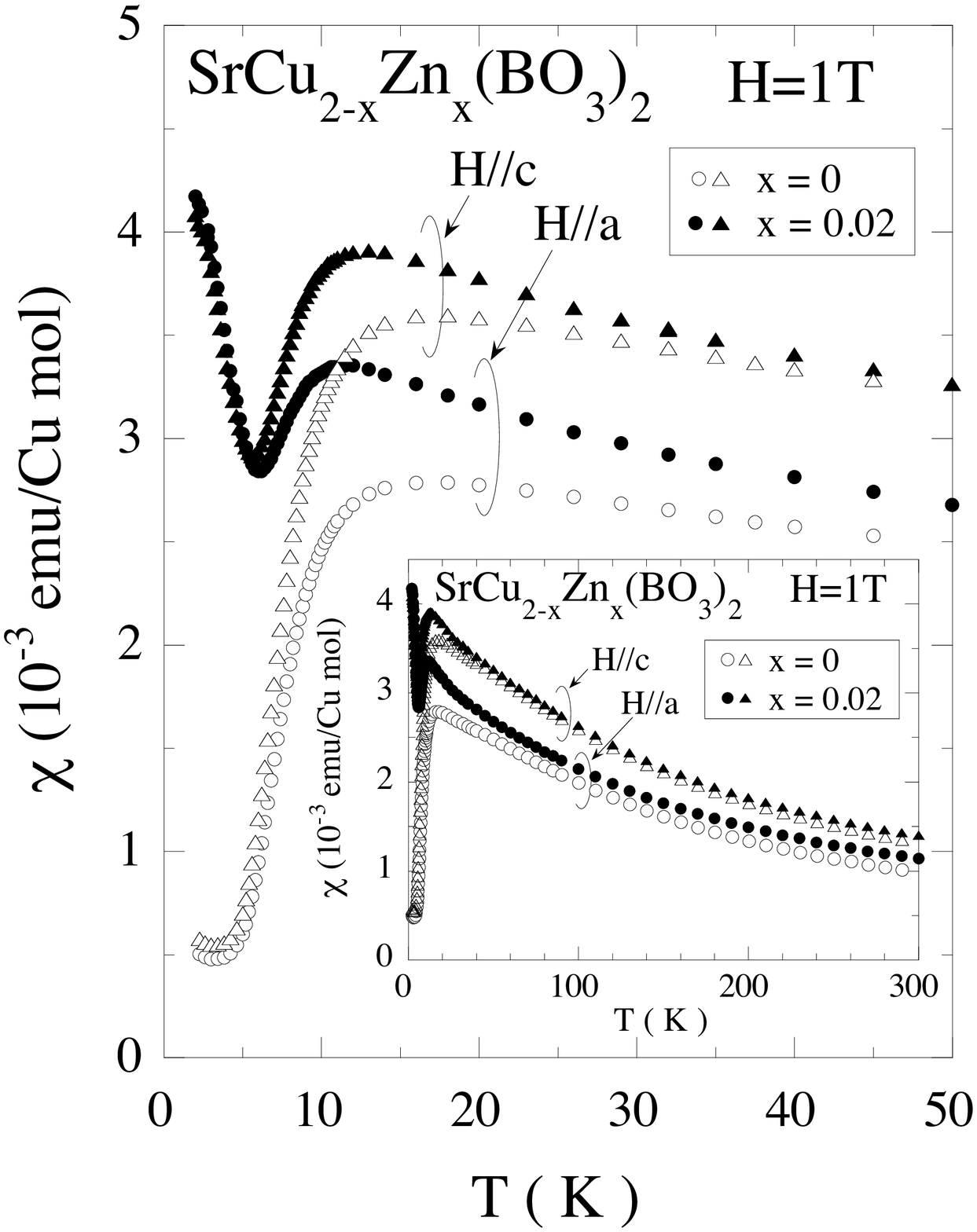}
\end{center}
\caption{Temperature dependence of the magnetic susceptibility $\chi$ in a magnetic field of 1 T parallel to the $a$- and $c$-axes for SrCu$_{2-x}$Zn$_x$(BO$_3$)$_2$ with $x =$ 0 and 0.02. 
The inset shows the data in a wide temperature-region.
}
\label{fig1}
\end{figure}

\begin{figure}[p]
\begin{center}
\includegraphics[width=0.5\linewidth]{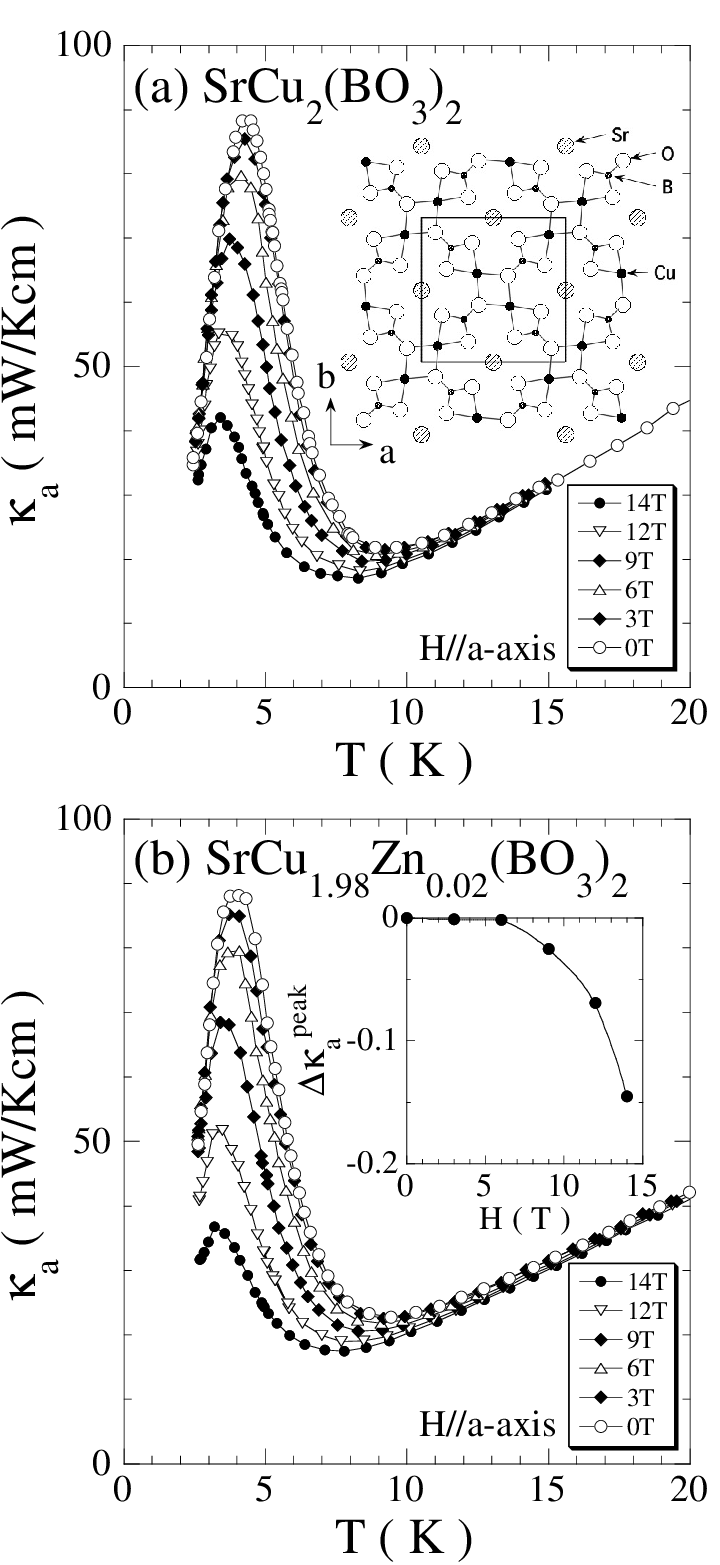}
\end{center}
\caption{Temperature dependence of the thermal conductivity parallel to the $a$-axis, $\kappa_{\rm a}$, of SrCu$_{2-x}$Zn$_x$(BO$_3$)$_2$ with (a) $x =$ 0 and (b) $x =$ 0.02. 
The inset of (a) is a schematic picture of the CuBO$_3$ plane in SrCu$_2$(BO$_3$)$_2$\cite{rf:Kageyama1}, where the solid square indicates the unit cell. 
The inset of (b) shows the magnetic field dependence of the relative reduction of the thermal conductivity peak due to the 1 \% Zn-substitution, $\Delta\kappa_{\rm a}^{\rm peak} = [\kappa_{\rm a}^{\rm peak}(x = 0.02) - \kappa_{\rm a}^{\rm peak}(x = 0)]/\kappa_{\rm a}^{\rm peak}(x = 0)$. }
\label{fig:kappa}
\end{figure}

Figure \ref{fig1} shows the temperature dependence of the magnetic susceptibility, $\chi$, for SrCu$_{2-x}$Zn$_x$(BO$_3$)$_2$. 
As reported in the previous report\cite{rf:Kageyama1,rf:KudoSrCu2}, the susceptibility of $x = 0$ exhibits a peak at 17 K and decreases with decreasing temperature at low temperatures below 17 K, independent of the applied field direction. 
This is a typical behavior reflecting the formation of spin-singlet dimers with a finite spin-gap. 
The small Curie-tail observed below 3.5 K may be due to magnetic impurities and/or defects of Cu$^{2+}$ ions. 
The peak temperature does not change through the 1\% Zn substitution, indicating no change in the spin gap.  
The relative large Curie-tail observed below 6 K may be due to isolated spins which are located adjacent to Zn and able to form no spin-singlet dimers.

Figure \ref{fig:kappa}(a) shows the temperature dependence of the thermal conductivity parallel to the $a$-axis, $\kappa_{\rm a}$, of SrCu$_2$(BO$_3$)$_2$ in magnetic fields. 
In zero field, $\kappa_{\rm a}$ exhibits an anomalous enhancement at low temperatures below $\sim$ 10 K, as reported in previous papers\cite{rf:KudoSrCu2,rf:HoffSrCu2,rf:KudoSrZn,rf:excuse}. 
The enhancement in zero field is attributed to the increase of $\kappa_{\rm phonon}$, which is caused by the suppression of phonon-magnon scattering owing to the spin-gap formation.
The suppression of the enhancement by the application of magnetic field is explained as being due to the reduction of the spin gap.
Figure \ref{fig:kappa}(b) shows the temperature dependence of $\kappa_{\rm a}$ in the 1\% Zn-substituted SrCu$_{1.98}$Zn$_{0.02}$(BO$_3$)$_2$ in magnetic fields. 
In low magnetic fields below 6 T, it appears that the peak of $\kappa_{\rm a}$ in the Zn-substituted sample coincides with that in the Zn-free one. 
This is because the enhancement of $\kappa_{\rm a}$ in the spin-gap state is due to the enhancement of $\kappa_{\rm phonon}$ and because the spin gap does not change through the 1\% Zn-substitution, as mentioned above. 
In high magnetic fields above 6 T, on the other hand, the peak of the Zn-substituted sample is smaller than that of the Zn-free one. 
For the clarity, the magnetic field dependence of the relative reduction of the thermal conductivity peak due to the 1\% Zn substitution, $\Delta\kappa_{\rm a}^{\rm peak} = [\kappa_{\rm a}^{\rm peak}(x = 0.02) - \kappa_{\rm a}^{\rm peak}(x = 0)]/\kappa_{\rm a}^{\rm peak}(x = 0)$, is plotted in the inset of Fig. \ref{fig:kappa}(b). 
It is found that $\Delta\kappa_{\rm a}^{\rm peak}$ decreases with increasing field at high fields above 6 T while it is negligibly small at low fields below 6 T. 
This result indicates that multi-triplet magnons contribute to the thermal conductivity above 6 T in the Zn-free sample more than in the Zn-substituted one. 
In fact, the excitation not only to the first-excited state of single-triplet magnons but also to the higher excited state of multi-triplet magnons has been pointed out to occur from the specific heat measurements\cite{rf:Amaya,rf:Alloy}. 
That is, the shottky-type peak splits into two peaks around 2--7 K in high fields above $\sim$ 12 T.
Therefore, it is concluded that the thermal conductivity peak in high fields above $\sim$ 6 T in the spin-gap state of SrCu$_2$(BO$_3$)$_2$ is composed of not only $\kappa_{\rm phonon}$ but also $\kappa_{\rm magnon}$ due to the multi-triplet magnons.

The thermal conductivity measurements were performed at the High Field Laboratory for Superconducting Materials, IMR, Tohoku University.
This work was supported by a Grant-in-Aid for Scientific Research of the Ministry of Education, Science, Sports and Culture, Japan, and also by CREST of Japan Science and Technology Corporation.



\begin{thebibliography}{99}
\bibitem{rf:excuse} The enhancement of $\kappa_a$ reported in Ref. \citen{rf:KudoSrCu2} is smaller than those in Refs. \citen{rf:HoffSrCu2,rf:KudoSrZn} and the present paper. 
This is because $\kappa_a$ in Ref. \citen{rf:KudoSrCu2} was measured by a two-terminal method where the thermal conductivity was underestimated due to the contact thermal resistance. 
In Refs. \citen{rf:HoffSrCu2,rf:KudoSrZn} and the present paper, meanwhile, $\kappa_{\rm a}$ was measured by a four-terminal method.
\end{thebibliography}
\end{document}